\documentclass[superscriptaddress,
 amsmath,amssymb,amsthm,mathrsfs,
 preprint,
 aps,
 prb,
 ]{revtex4-1}

\usepackage{graphicx}
\usepackage{dcolumn}
\usepackage{bm}

\usepackage[
margin=0.75in,
]{geometry}
\usepackage{color}
\usepackage{multirow}
\DeclareMathAlphabet{\mathpzc}{OT1}{pzc}{m}{it}

\def\dr{$d(r)$}
\def\R3m{$R\overline{3}m$}
\def\C2m{$C2/m$}
\def\C2{$C2$}
\newcommand{\ra}[1]{\renewcommand{\arraystretch}{#1}}

\begin{document}                  



\title{
Magnetic structure determination from the magnetic pair distribution function (mPDF): ground state of MnO}

\author{Benjamin A. Frandsen}
\affiliation{%
 Department of Physics, Columbia University, New York, NY 10027, USA.
}%

\author{Simon J. L. Billinge}
\email{sb2896@columbia.edu}
\affiliation{%
 Condensed Matter Physics and Materials Science Department, Brookhaven National Laboratory, Upton, NY 11973, USA.
}%
\affiliation{%
 Department of Applied Physics and Applied Mathematics, Columbia University, New York, NY 10027, USA.
}%

\begin{abstract}
An experimental determination of the magnetic pair distribution function (mPDF) defined in an earlier paper
(Frandsen, Yang, \& Billinge. (2014) \textit{Acta Crystallogr. A}, \textbf{70}(1), 3-11) is presented for the first time. The mPDF was determined from neutron powder diffraction data from a reactor and a neutron time-of-flight total scattering source on a powder sample of the antiferromagnetic oxide MnO.
A description of the data treatment that allowed the measured mPDF to be extracted and then modelled is provided and utilized to investigate the low-temperature structure of MnO. Atomic and magnetic co-refinements support the scenario of a locally monoclinic ground-state atomic structure, despite the average structure being rhombohedral, with the magnetic PDF analysis successfully recovering the known antiferromagnetic spin configuration. The total scattering data suggest a preference for the spin axis to lie along the pseudocubic [10$\bar{1}$] direction. Finally, $r$-dependent PDF refinements indicate that the local monoclinic structure tends toward the average rhombohedral \R3m\ symmetry over a length scale of approximately 100~\AA. 
\end{abstract}

\maketitle


\section{Introduction}
In a recent paper, the theoretical equations for the magnetic pair distribution function (mPDF) were derived~\cite{frand;aca14}. The mPDF, which can be obtained experimentally by properly normalizing and Fourier transforming the magnetic scattering intensity from a neutron powder diffraction measurement in similar fashion to the atomic PDF~\cite{egami;b;utbp12}, is an intuitive function that reveals local magnetic correlations directly in real space. Since the mPDF technique places diffuse and Bragg scattering on an equal footing, it provides sensitivity to both short- and long-range magnetic order. The intuitive nature of the mPDF allows one to gain
useful information about magnetic correlations by direct inspection of real-space mPDF patterns, which is often difficult when magnetic scattering
is viewed only in reciprocal space. In addition, the mPDF can be calculated easily and quickly for a given magnetic structure, allowing for rapid quantitative refinement of magnetic structural models.

\begin{figure}
\includegraphics[width=8.85cm]{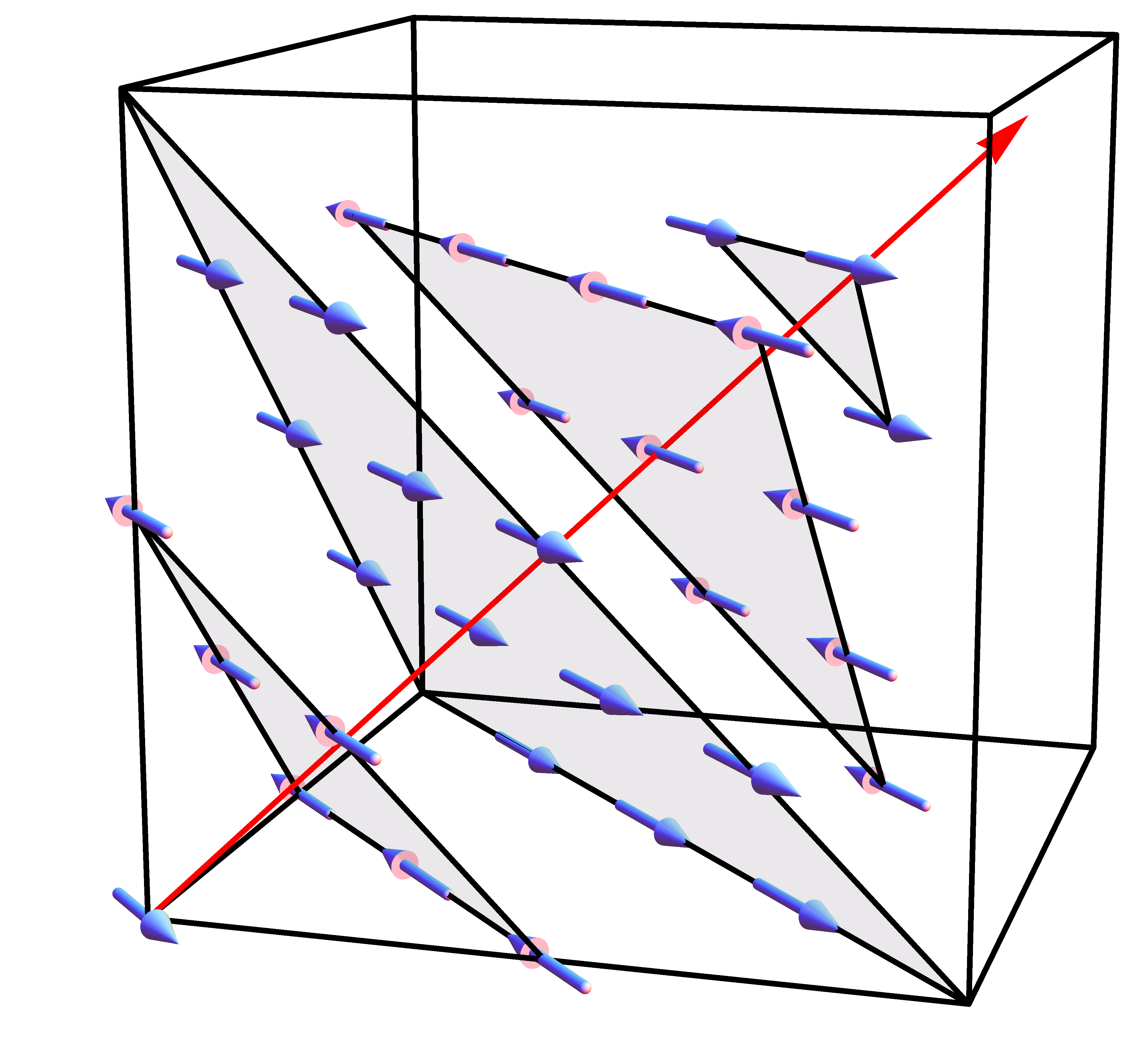}
\caption{Antiferromagnetic structure of MnO. The spin alignment axis lies within the (111) plane and reverses between adjacent sheets along the [111] direction.}
\label{fig;MnOspins}
\end{figure}

Here we report the first experimental demonstration of the mPDF technique, which we use to investigate the magnetic ground state of MnO. Early neutron diffraction studies~\cite{shull;pr51,roth;pr58} showed that MnO has the cubic rock-salt structure at
high temperature and undergoes an antiferromagnetic transition with a Neel temperature of $T_{\mathrm{N}}=118$~K. This magnetic transition is accompanied by a rhombohedral compression of the lattice along the [111] direction, resulting in \R3m\ symmetry~\cite{roth;pr58}. The spins of the Mn$^{2+}$ ions lying within common (111) planes align ferromagnetically, with antiferromagnetic alignment between adjacent sheets along the [111] stacking direction resulting in the so-called type-II antiferromagnetic structure (see Fig.~\ref{fig;MnOspins}). The spin alignment axis is known to lie within the (111) plane. However, for a structure in the \R3m symmetry space group, the absolute orientation of the spin within the plane
cannot be determined from powder diffraction measurements~\cite{goodw;prl06}.

Paradoxically, it has been known for some time that the antiferromagnetic spin arrangement in MnO is compatible only with monoclinic or lower symmetry~\cite{shake;prb88}, so the true structural symmetry must be lower than \R3m. However, not even high-resolution neutron diffraction measurements revealed a deviation from \R3m\ symmetry in the average structure~\cite{shake;prb88}, which presents a conundrum.  One possibility is that the local structure has the required monoclinic or lower symmetry, but averaging over differently ordered domains on longer length scales results in the higher symmetry average structure. 

Neutron total scattering techniques can often provide sensitivity to these types of local distortions. The first total scattering study of MnO~\cite{melle;jpcm98} used the reverse Monte Carlo (RMC) approach to investigate the nuclear and magnetic structure at various temperatures. The RMC algorithm is typically used to perform structural refinements in reciprocal space, in contrast to the real-space approach of PDF. Descriptions of RMC modelling and examples of its application to atomic and magnetic structure can be found elsewhere~\cite{mcgre;jpcm01,goodw;prl12}. In this first total-scattering study of MnO, the low-temperature data were found to be consistent with the rhombohedral nuclear structure, although no specific attempts to investigate the possibility of local monoclinicity were reported. On the other hand, more recent RMC refinements did find evidence for a local monoclinic symmetry over a length scale of at least 20~\AA~\cite{goodw;prl06}, limited by the size of the RMC box. The RMC configurations projected into a monoclinic unit cell in which the atomic positions were displaced very slightly from their high-symmetry sites and a small out-of-plane spin component was found to vary sinusoidally along the pseudocubic [111] direction within the unit cell, resulting in a local $C2$ symmetry. In this lower symmetry space group, it is possible, in principle, to measure the absolute spin orientation within the pseudocubic (111) plane even in the powder-averaged scattering pattern. Exploiting this sensitivity, the RMC analysis found a very slight preference for the spin axis to lie along the pseudocubic $\langle 11\bar{2} \rangle$ directions.

Here we revisited this problem by measuring and analyzing the mPDF from MnO at low temperature. This serves to assess the utility of this new approach, as well as to verify different aspects of the earlier studies.  We find that a monoclinic model with the known type-II antiferromagnetic ordering gives a better fit than the rhombohedral model to both the atomic and magnetic PDF data in the low-$r$ region, supporting the scenario of a local monoclinic distortion, though we find smaller atomic displacements than the earlier study \cite{goodw;prl06}. Fitting the data in real space allowed us to investigate the $r$-dependence of the atomic and magnetic PDF fits over a wide range of $r$, which revealed that the proposed local monoclinicity of the structure appears to weaken with increasing $r$ and crosses over to a rhombohedral structure on a length scale of $\sim$ 10~nm.  Our analysis reveals a tendency for the spins to align along the pseudocubic $[10\bar{1}]$ directions, but does not provide evidence for the sinusoidal arrangement of canted spins.  Finally, we tested the ability of solving a magnetic structure from mPDF data by allowing unconstrained mPDF fits, where all the spins within the monoclinic unit cell were given random initial orientations and allowed to vary independently.  This successfully recovered the expected antiferromagnetic structure.  This work demonstrates the utility and intuitive power of the mPDF approach for studying magnetism in materials.

\section{Theory}
To review, the mPDF for a system with a single spin-only magnetic species is given by~\cite{frand;aca14}
\begin{align}
\mathpzc{f}(r)&=\frac{2}{\pi}\int_{0}^{\infty} Q\left(\frac{I_m}{\frac{2}{3}N_s S(S+1)(\gamma r_0)^2 f_{m}^{2}(Q)}-1\right)\sin{Q r}\text{d}Q \label{FT}
\\&=\frac{1}{N_s}\frac{3}{2S(S+1)}\sum\limits_{i \ne j}\left( \frac{A_{ij}}{r}\delta (r-r_{ij})+B_{ij}\frac{r}{r_{ij}^3}\Theta (r_{ij}-r)\right) \label{fullfofr},
\end{align}
where $I_m$ is the orientationally averaged magnetic scattering cross section~\cite{blech;p64}, the subscripts $i$ and $j$ refer to individual magnetic moments~$ \boldsymbol{S_{\textit i}}$ and~$\boldsymbol{S_{\textit j}}$ separated by a distance~$r_{ij}$, $A_{ij}=\langle S^y_i S^y_j \rangle$, $B_{ij}=2\langle S^x_i S^x_j \rangle - \langle S^y_i S^y_j \rangle$,  $S$ is the spin quantum number in units of~$\hbar$, $r_0=\frac{\mu _0}{4\pi}\frac{e^2}{m_e}$ is the classical electron radius, $\gamma = 1.913$ is the neutron magnetic moment in units of nuclear magnetons, $f_{m}(Q)$ is the magnetic form factor, $N_s$ is the number of spins in the system, and~$Q$ is the magnitude of the scattering vector. The coordinate system used to express~$A_{ij}$ and~$B_{ij}$ is locally defined for each spin pair such that $\hat{\boldsymbol{x}}$ lies along the vector joining the pair of spins (pair axis) and $\hat{\boldsymbol{y}}$ is chosen such that the $xy$-plane contains the pair axis and the first spin, i.e.  $\hat{\boldsymbol{x}}=\frac{\boldsymbol{r_{\textit j}}-\boldsymbol{r_{\textit i}}}{\vert \boldsymbol{r_{\textit j}}-\boldsymbol{r_{\textit i}} \vert}$ and $\hat{\boldsymbol{y}}=\frac{\boldsymbol{S_{\textit i}}-\hat{\boldsymbol{x}}(\boldsymbol{S_{\textit i}}\cdot \hat{\boldsymbol{x}})}{\vert\boldsymbol{S_{\textit i}}-\hat{\boldsymbol{x}}(\boldsymbol{S_{\textit i}}\cdot \hat{\boldsymbol{x}})\vert}$. We have chosen to denote the mPDF as $\mathpzc{f}(r)$ in accordance with the notation introduced by Farrow and Billinge~\cite{farro;aca09}, where $\mathpzc{f}(r)$ is the full Fourier transform of the properly normalized intensity from $Q=0$ to $Q=\infty$. We note that this is not strictly equivalent to the function $G(r)=4\pi r(\rho(r)-\rho_0)$ that is commonly used in structural PDF.

The term in Eq.~\ref{fullfofr} containing the delta function depends on the spin correlations transverse to the axis joining
the spin pair, and is positive for ferromagnetic and negative for antiferromagnetic transverse correlations.
The second term is linear in $r$ with a slope that is dominated by the longitudinal correlations,
again with a positive (negative) slope generally indicating ferromagnetic (antiferromagnetic) correlations~\cite{frand;aca14}.
Although the mPDF was calculated for a number of magnetic structures in the previous work, an experimental verification
has not yet been demonstrated.

\section{Methods}

\subsection{Neutron scattering measurements}
A large powder sample of MnO (99\% pure) was obtained from a commerical supplier (Alfa Aesar). We performed neutron scattering measurements on approximately 5~g of MnO at 300~K and 5~K on beamline D20 at the Institut Laue-Langevin (ILL) using an incident wavelength of 0.94079~\AA, resulting in a maximum momentum transfer of $Q=12.9$~\AA$^{-1}$. The background signal from the vanadium sample can and cryostat was measured separately at each temperature and subtracted from the observed scattering patterns. The presence of long-range magnetic order was confirmed by the appearance of additional Bragg peaks at low temperature arising from the antiferromagnetic spin structure.

We also performed neutron time-of-flight total scattering measurements on the NPDF instrument at
the Lujan Neutron Scattering Center at Los Alamos National Laboratory. These
measurements were optimized for structural PDF refinement and accordingly benefitted
from very low backgrounds and wide momentum-transfer coverage. Standard corrections were applied to all data sets.


\subsection{mPDF data processing}
The most direct method of obtaining the mPDF $\mathpzc{f}(r)$ experimentally is to
isolate the magnetic scattering intensity $I_m$, normalize by dividing
by the square of the magnetic form factor, and compute the Fourier transform. The measured scattered intensity
obtained from a typical experiment using unpolarized neutrons contains both the nuclear scattering $I_n$ as well as $I_m$. Therefore, one
must find an accurate means of separating the two contributions. In the following, we describe two approaches we have taken to address this problem: first, we separate the nuclear and magnetic contributions in reciprocal space; and second, we separate them in real space.

\subsubsection{Reciprocal-space separation of nuclear and magnetic signals}
\begin{figure}
\includegraphics[width=6cm]{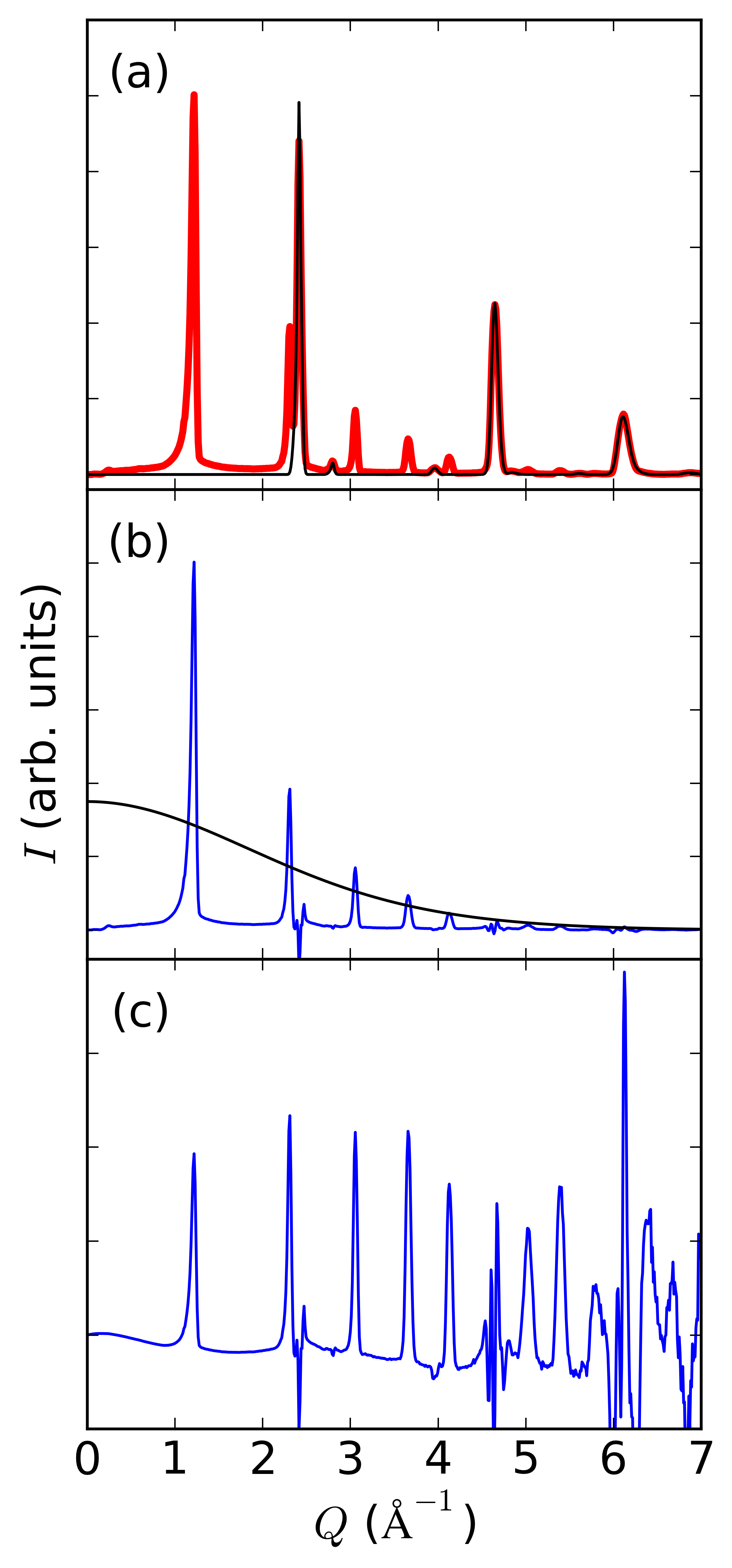}
\caption{Magnetic scattering from MnO at 5~K. (a) Rietveld refinement of the \R3m\ structure against the nuclear Bragg peaks. The red curve is the observed scattering pattern, black is the calculated pattern. (b) Experimental scattering pattern with the nuclear fit subtracted, consisting of the magnetic scattering and residual errors in the nuclear fit, with the square of the magnetic form factor $f_{m}^2 (Q)$ overlaid. (c) Experimental scattering pattern after normalization by $f_{m}^2 (Q)$. Large errors are seen at high $Q$. }
\label{fig;ILL-panel-fitsub}
\end{figure}

If the nuclear structure is known and well ordered, then a fit can be performed to the nuclear
Bragg peaks and subsequently they can be subtracted from the original scattering
pattern. The remaining signal, which contains $I_m$ and any residual errors from the nuclear fit, can then be normalized and Fourier transformed to obtain the mPDF. We applied this strategy to the 5~K scattering pattern measured at ILL. We first subtracted the sample-environment background signal from the raw data and then used the Rietveld refinement program FullProf~\cite{rodri;unpub90} to fit the rhombohedral model to the nuclear Bragg peaks, without refining any additional background contribution. We then subtracted the fitted nuclear profile and normalized the remaining signal by the square of the magnetic form factor, $f_{m}^{2}(Q)$, to obtain the quantity that is Fourier transformed in Eq.~\ref{FT}. This is illustrated in Fig.~\ref{fig;ILL-panel-fitsub}. Finally, we numerically computed the Fourier transform up to $Q_{\mathrm{max}}=7$~\AA$^{-1}$ to obtain the experimental mPDF. Throughout this work, we use the analytic approximation for the Mn$^{2+}$ magnetic form factor provided in the International Tables~\cite{wilso;b;itc95}.

\begin{figure}
\includegraphics[width=6cm]{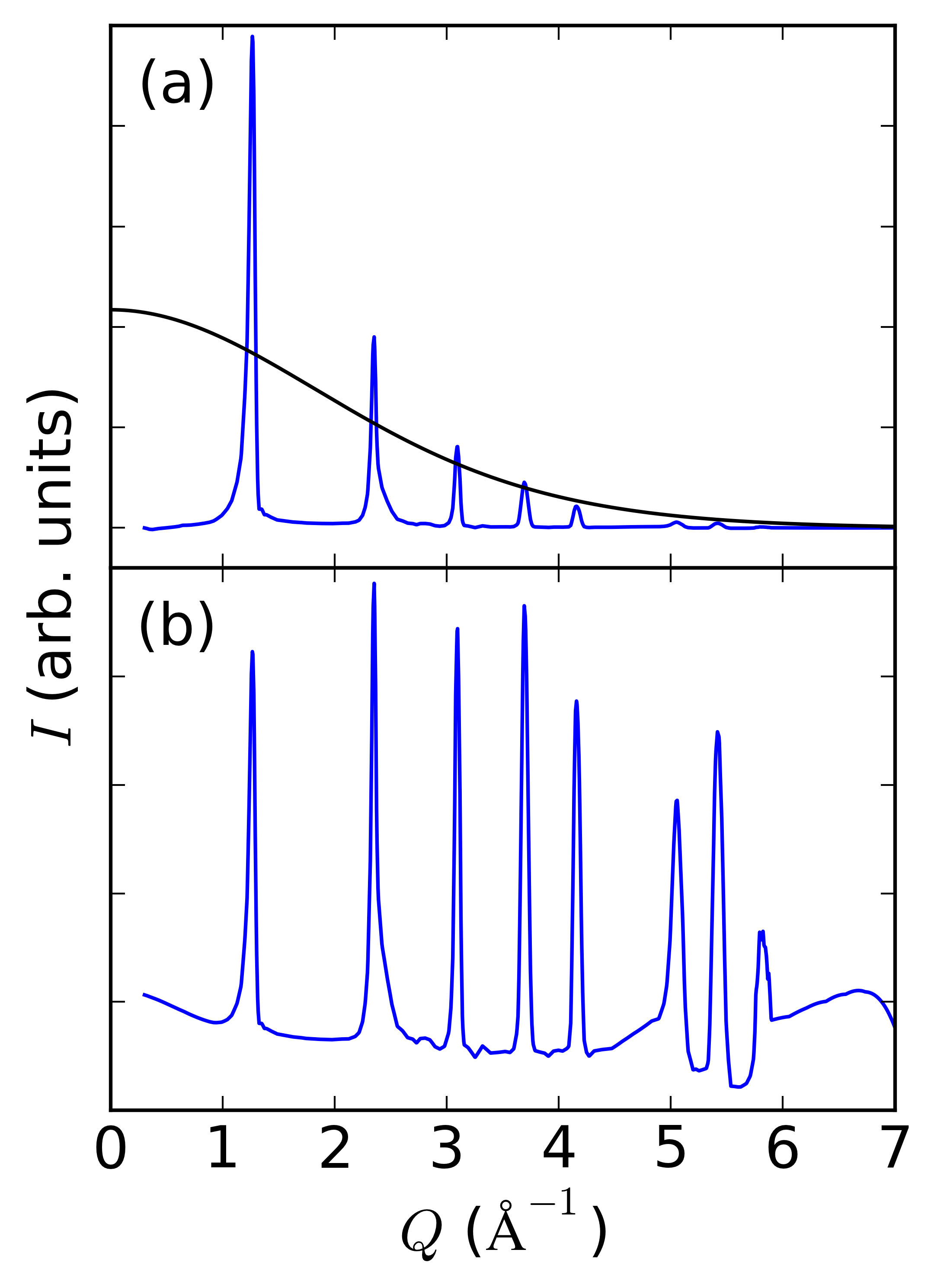}
\caption{Corrected magnetic scattering intensity from MnO. (a) Magnetic scattering at 5~K obtained by subtracting the fitted nuclear Bragg peaks and manually correcting the obvious errors, overlaid by the square of the magnetic form factor $f_{m}^2 (Q)$. (b) Corrected magnetic scattering after normalization by $f_{m}^2 (Q)$.}
\label{fig;ILL-panel-handpicked}
\end{figure}

It is clear that the preceding procedure resulted in significant interference from errors in the nuclear fit, particularly at high $Q$ where the small value of $f_{m}(Q)$ amplifies the errors during the normalization step. We attempted to correct this by manually smoothing the magnetic scattering pattern to remove the most obvious errors after subtraction of the fitted nuclear scattering. We then normalized and Fourier transformed the hand-corrected data in the same manner as before. This is displayed in Fig.~\ref{fig;ILL-panel-handpicked}.
We note that the use of polarized neutrons with a powder-optimized measurement scheme such
as the $xyz$-polarization~\cite{stewa;jac09,schwe;jpconfs10} or 10-point method~\cite{ehler;rsi13} would automatically separate the magnetic and nuclear scattering contributions and could be very effective for mPDF measurements, as long as the scattered intensity can be measured up to a sufficient momentum transfer ($Q\gtrsim 7$~\AA$^{-1}$ for typical transition-metal magnetic systems).  However, this is beyond the scope of the current paper.

\subsubsection{Real-space separation of nuclear and magnetic signals}
The difficulties of separating the nuclear and magnetic signals in reciprocal space motivated us to seek an improved means of obtaining reliable experimental mPDFs. Using the total scattering data obtained on the NPDF instrument, we developed an approach to obtain the mPDF simultaneously
with the atomic PDF by Fourier transforming both the nuclear and magnetic scattering together and separating their respective signals in real space. This opens the possibility for structural and magnetic PDF co-refinement.

\begin{figure}
\includegraphics[width=8.85cm]{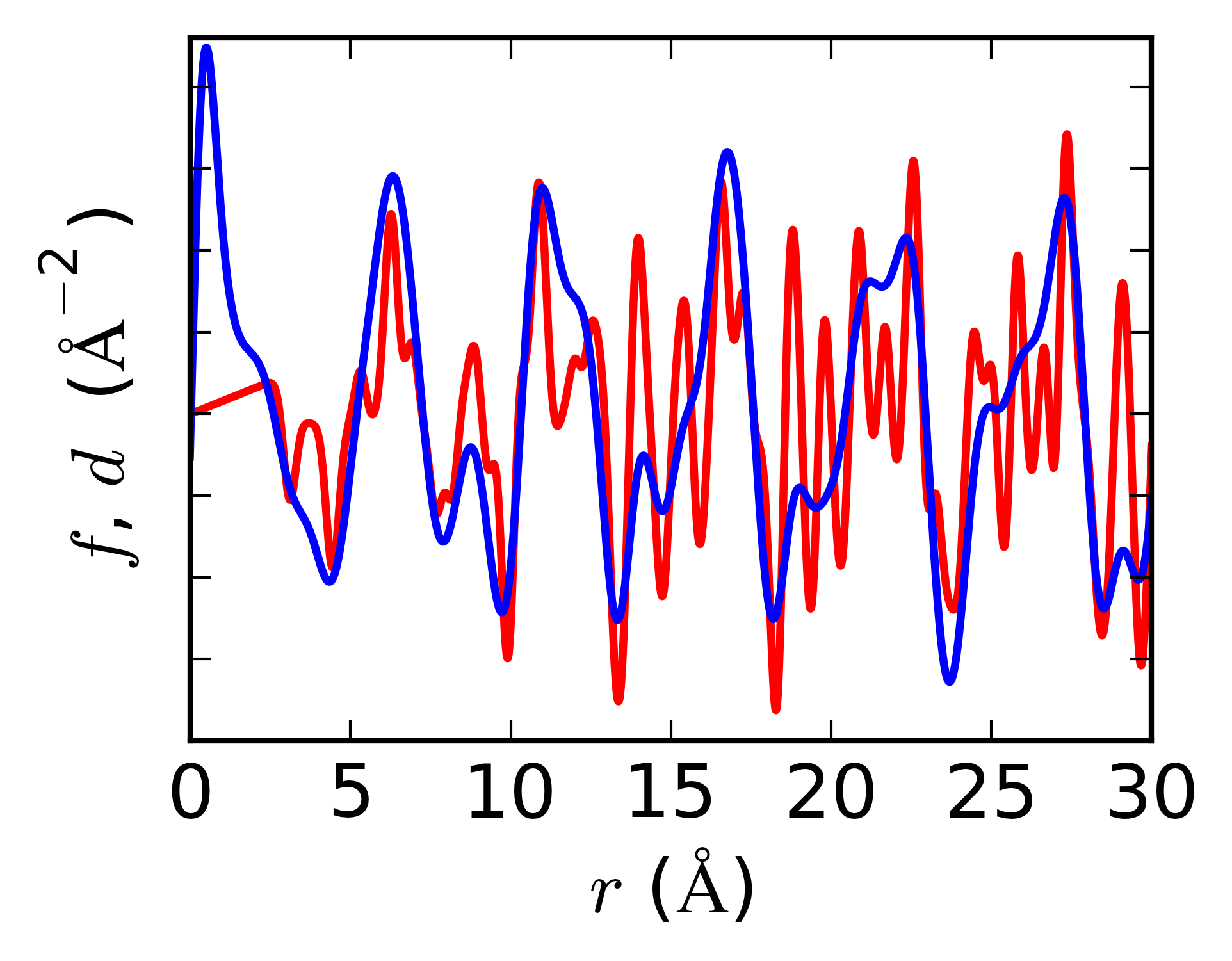}
\caption{Comparison of the properly normalized mPDF, $\mathpzc{f}(r)$ (red), and the unnormalized mPDF, $d(r)$ (blue), calculated for MnO. Details are provided in the text.}
\label{fig;frdrcomp}
\end{figure}

According to standard structural PDF protocols, for example as implemented by the software program PDFgetN~\cite{peter;jac00}, the PDF is obtained from the total scattering intensity as
\begin{align}
G_{tot}(r)=\mathcal{F}\left\{ Q\left(\frac{I_{tot}}{N_a\langle b\rangle ^2}-\frac{\langle b^2\rangle}{\langle b \rangle^2}\right)\right\} ,
\end{align}
where $\mathcal{F}\left\{ \cdots \right\} $ is shorthand for the Fourier transform and includes a constant prefactor, $I_{tot}=I_n+I_m$ is the total scattered intensity including the nuclear contribution $I_n$ and the magnetic contribution $I_m$, $N_a$ is the number of atoms, and $b$ is the nuclear scattering length, with angled brackets denoting an average over all nuclei present in the sample. Separating the nuclear and magnetic contributions yields
\begin{align}
G_{tot}(r)&=\mathcal{F}\left\{ Q\left(\frac{I_{n}}{N_a\langle b\rangle ^2}-\frac{\langle b^2\rangle}{\langle b \rangle^2}\right)\right\} +\mathcal{F}\left\{ Q\frac{I_{m}}{N_a\langle b\rangle ^2}\right\}
\\&=G_{n}(r)+d(r)/N_a\langle b \rangle ^2,\label{totalPDF}
\end{align}
where $G_{n}(r)$ is the atomic (nuclear) PDF and $d(r)=\mathcal{F}\left\{ Q I_m(Q)\right\} $ is a quantity that we will call the ``unnormalized mPDF,'' since it does not involve division by the magnetic form factor $f_{m}(Q)$. A straightforward application of the convolution theorem reveals that
\begin{align}
d(r)=C_1 \times \mathpzc{f}(r)\ast S(r) + C_2 \times \frac{\textrm{d}S}{\textrm{d}r},\label{eq;dr}
\end{align}
where $C_1$ and $C_2$ are constants related by $C_1 / C_2 = -1 / \sqrt{2\pi}$ in the fully ordered state, $\ast$ represents the convolution operation, and $S(r)=\mathcal{F}\left\{f_{m}(Q)\right\}\ast \mathcal{F}\left\{ f_{m}(Q)\right\}$. The quantity $\mathcal{F}\left\{f_{m}(Q)\right\}$ is closely related to the real-space spin density. Roughly speaking, \dr\ is equivalent to the proper mPDF $\mathpzc{f}(r)$ twice broadened by the spin density with an additional peak at low $r$ produced by the derivative term in Eq.~\ref{eq;dr}. The two functions $\mathpzc{f}(r)$ and \dr\ are illustrated by calculating them for for the rhombohedral model of MnO in Fig.~\ref{fig;frdrcomp}. Although the features in \dr\ are significantly broader than those in $\mathpzc{f}(r)$, the overall antiferromagnetic structure is still clearly evident in \dr, and it has the advantage of being obtained directly from the Fourier transformed data. Throughout the remainder of this paper, the term ``mPDF'' may be applied to either $\mathpzc{f}(r)$ or \dr,
but the context will make it clear which function is meant. More details about the derivation of \dr\ are provided in the appendix.
Using the program PDFgetN, we applied standard background corrections to the total scattering data and generated the experimental combined atomic and magnetic PDF by Fourier transforming the total scattering intensity up to  $Q_{\mathrm{max}}=35$~\AA$^{-1}$. The PDF refinement program PDFgui~\cite{farro;jpcm07} was used to perform atomic PDF refinements of the rhombohedral \R3m model and the monoclinic $C2$ model reported in the previous RMC study~\cite{goodw;prl06}. After the atomic structure refinement, the magnetic structure was generated with the Mn spins fixed to the Mn ions according to the refined structural parameters of the atomic PDF fits, and mPDF refinements were performed against the difference curve, which predominantly consists of the unnormalized mPDF $d(r)$. For reference, the monoclinic lattice vectors $\mathbf{a'}$, $\mathbf{b'}$, and $\mathbf{c'}$ are related to the high-temperature cubic lattice vectors $\mathbf{a}$, $\mathbf{b}$, and $\mathbf{c}$ through the transformation
\begin{align}
\begin{bmatrix} \mathbf{a'} \\ \mathbf{b'} \\ \mathbf{c'} \end{bmatrix} = \left[\begin{array}{ccc} \tfrac{1}{2} & -1 & \tfrac{1}{2} \\ \tfrac{1}{2} & 0 & -\tfrac{1}{2} \\ 2 & 2 & 2 \end{array}\right]  \begin{bmatrix} \mathbf{a} \\ \mathbf{b} \\ \mathbf{c} \end{bmatrix}.
\end{align}

The mPDF refinements were conducted with a custom-built program written in the Python programming language. The refinable parameters were the spin orientations, the two scaling constants in Eq.~\ref{eq;dr}, an overall Gaussian damping factor to represent the finite correlation length of the magnetic structure, and a Gaussian broadening factor to represent thermal motion. We note that there are two distinct thermal effects that can broaden the mPDF: motion of the ionic sites on which the spins reside, and fluctuations of the relative orientations of the spins. The best practice would be to treat these two cases separately, extracting the appropriate distribution of pair distances from the atomic PDF analysis and then refining only the broadening contribution that arises from thermal motion of the spin orientations. However, as a first attempt we use a single Gaussian broadening factor to account for both of these effects.  The effects of the finite experimental values of $Q_{\mathrm{min}}$ and $Q_{\mathrm{max}}$ were modelled by convoluting the calculated mPDF with the termination function~\cite{peter;jac03} $\theta(r;Q_{\mathrm{min}},Q_{\mathrm{max}})=\frac{Q_{\mathrm{max}}}{\pi}j_{0}(Q_{\mathrm{max}}r)-\frac{Q_{\mathrm{min}}}{\pi}j_{0}(Q_{\mathrm{min}}r)$, where $j_0$ represents the zeroth-order spherical Bessel function.

Two types of mPDF fits were performed: a local search (refinement) in which a single global spin axis was refined and the overall antiferromagnetic structure was imposed as a constraint, and a global optimization (structure solution) in which the orientations of each of the 12 spins in the monoclinic unit cell were given random initial starting directions and allowed to vary independently. A typical fit over a range of 100~\AA\ required a computation time on the order of seconds on a standard laptop for the constrained fits and minutes for the unconstrained fits, relatively fast compared to more computationally intensive methods such as RMC.

Finally, we implemented an iterative atomic/magnetic PDF fitting procedure as follows. After the initial atomic and magnetic refinements, we subtracted the calculated
\dr\ from the original real-space data to isolate the structural contribution,
refined the atomic PDF against this new data set, generated a new difference curve from
this refined structural model, used this to refine the magnetic structure again, and so on.

\section{Results and Discussion}

\subsection{mPDF analysis after reciprocal-space separation of nuclear and magnetic signals}
The result of a Rietveld refinement of the rhombohedral model of MnO against the nuclear Bragg peaks measured at the ILL is shown in Fig.~\ref{fig;ILL-panel-fitsub}(a), with the residual intensity after subtraction of the fitted nuclear profile shown in panel (b). This residual intensity contains the magnetic scattering and any errors in the nuclear fit. The square of the magnetic form factor, $f_{m}^{2}(Q)$, is also displayed in panel (b). Finally, the quantity that is Fourier transformed in Eq.~\ref{FT}, which we call $F_m(Q)$ and which is obtained by normalizing magnetic scattering $I_m$ by $f_{m}^{2}(Q)$, is given in panel (c).

\begin{figure}
\includegraphics[width=8.85cm]{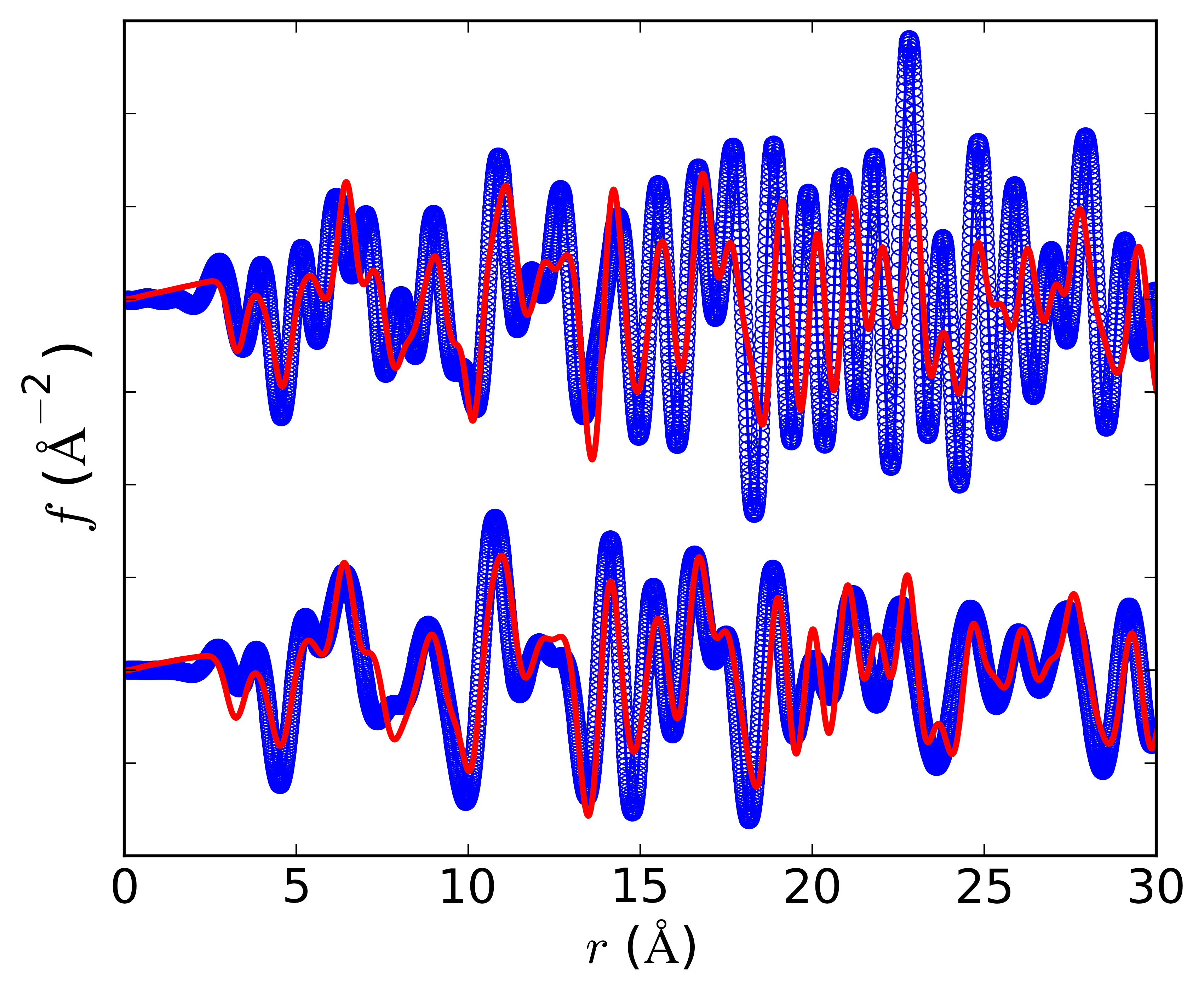}
\caption{mPDF of MnO obtained by Fourier transforming the normalized scattered intensity after subtraction of the fitted nuclear Bragg peaks (top) and after applying manual corrections to the nuclear fit before subtracting (bottom). Blue circles represent the experimental data and the red curves represent the fitted mPDFs.}
\label{fig;D20-fr-both}
\end{figure}

It is immediately apparent that the errors in the nuclear fit, which are reasonably small on the scale of the scattered intensity as seen in panels (a) and (b), become problematic at high $Q$ after normalization due to the exceedingly small value of $f_{m}^{2}(Q)$ above $\sim 4.5$~\AA$^{-1}$. The Fourier transform of $F_m(Q)$ up to $Q_{\mathrm{max}}=7$~\AA$^{-1}$ is displayed as the top curve in Fig.~\ref{fig;D20-fr-both}, overlaid by the calculated mPDF for the known type-II antiferromagnetic structure of MnO. Despite the aberrations introduced by the subtraction of the nuclear Bragg peaks, reasonable semi-quantitative agreement exists between the experimental and calculated patterns, in the sense that the peak positions are well reproduced, albeit with significant errors in the intensity.

Substantial improvements are obtained by manually removing the most obvious errors after subtraction of the fitted nuclear scattering, thereby obtaining a closer approximation to the actual magnetic scattering for subsequent normalization and Fourier transformation. Fig.~\ref{fig;ILL-panel-handpicked}(a) displays the magnetic scattering after performing this type of manual correction, and panel (b) shows the normalized scattering $F_m(Q)$.
The effects of these corrections on the mPDF can be seen in the bottom curve of Fig.~\ref{fig;D20-fr-both}, where the agreement between the experimental and calculated mPDFs is significantly better than in the previous case. These results constitute the first experimental verification of the mPDF expression given in Eq.~\ref{fullfofr}. The antiferromagnetic structure of MnO is intuitively captured by the alternating negative and positive peaks of the mPDF.

\subsection{Real-space atomic and magnetic PDF co-refinement}
\begin{figure}
\includegraphics[width=8.85cm]{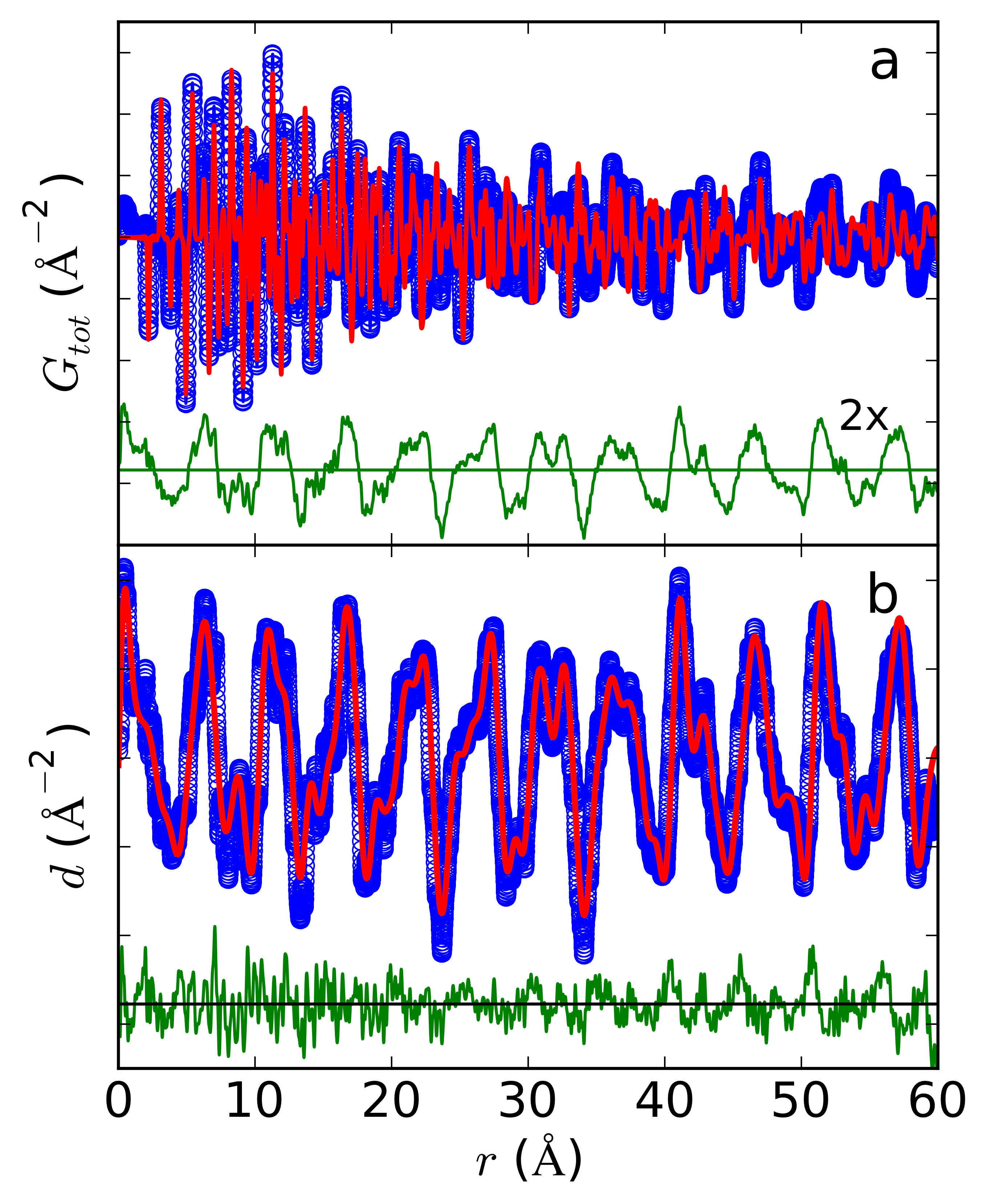}
\caption{(a) Atomic PDF refinement of the rhombohedral model for the nuclear structure of MnO. The mPDF $d(r)$ is seen in the difference curve, which is multiplied by two for clarity. (b) Refinement of the mPDF function $d(r)$ against the experimental mPDF obtained after subtraction of the refined atomic PDF.  In both panels, the blue curve is the experimental data, the red curve is the calculated pattern, and the offset green curve is the difference.}
\label{fig;rhombofits}
\end{figure}

We now move to the analysis of the total-scattering data collected on the NPDF instrument. After generating the Fourier transform of the combined nuclear and magnetic signals, we performed atomic PDF refinements, using first the rhombohedral model and then the monoclinic $C2$ model reported in the previous RMC study~\cite{goodw;prl06}. The result of the rhombohedral refinement over 60~\AA\ is shown in Fig.~\ref{fig;rhombofits}(a). This model adequately captures the sharp peaks of the atomic PDF but is of course unable to capture the unnormalized mPDF \dr, resulting in a large difference curve consisting of \dr\ and any residual errors in the fit to the atomic structure. The features in the difference curve reflect the mPDF from the antiferromagnetic structure of MnO.

We next attempted an mPDF fit to the experimental \dr\ contained in the atomic PDF difference curve. We enforced the overall antiferromagnetic spin configuration as a constraint but allowed the direction of the spin axis to vary freely from a randomly selected initial direction. The only additional parameters we refined were the scaling constants $C_1$ and $C_2$ and a Gaussian broadening factor. A view of an mPDF refinement from 0-60~\AA\ is provided in Fig.~\ref{fig;rhombofits}(b). The calculated mPDF from the model clearly does an excellent job of fitting the measured mPDF signal over the entire range, with just a small, oscillatory residual signal. At first glance it may appear that only one or two dominant frequencies are necessary to capture the overall shape of the mPDF, and so one might ask whether a fit to the mPDF would be unique and contain enough detail to be useful. However, closer inspection reveals that the actual structure of the mPDF is more complex, with various split peaks, side peaks, and long-range modulations that are quite sensitive to the details of the magnetic structure. The ability to perform reliable fits with the experimental mPDF is borne out by the robust convergence of the spin axis from a randomly generated starting orientation to a direction in the (111) plane, in agreement with the known spin axis. These results verify the expression for the unnormalized mPDF \dr\ given in Eq.~\ref{eq;dr}.

\begin{figure}
\includegraphics[width=18.0cm]{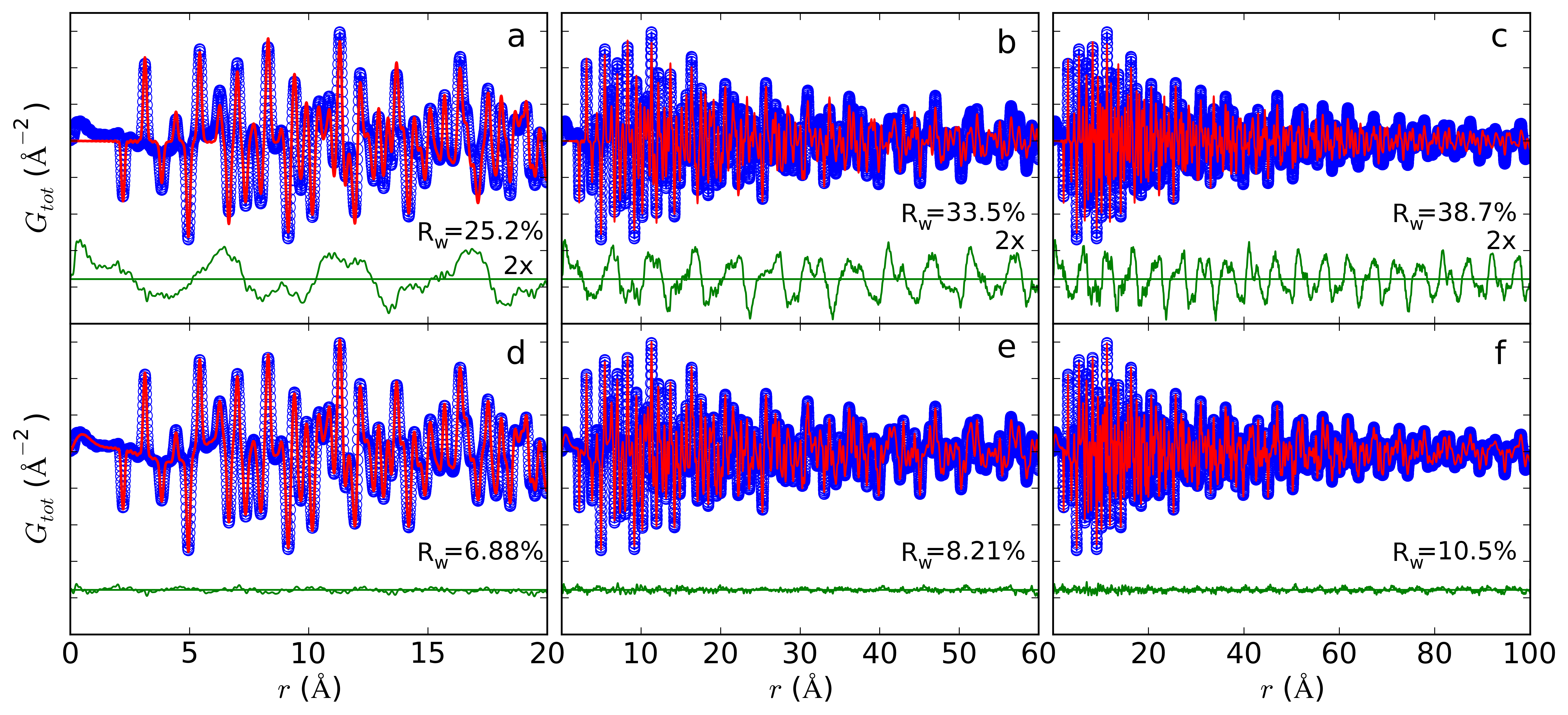}
\caption{Atomic and magnetic PDF refinements of the monoclinic model for three fitting ranges. (a)-(c) Refinement results when only the atomic structure is included in the model. The mPDF $d(r)$ is seen in the difference curves, which are multiplied by two for clarity. (d)-(f) Results of co-refinements of the atomic and magnetic PDFs. In all panels, the blue curve is the experimental data, the red curve is the calculated pattern, and the offset green curve is the difference.}
\label{fig;combofits}
\end{figure}

We now present a more detailed analysis using the proposed monoclinic structure. The results of atomic PDF refinements for various fitting ranges are shown in Fig.~\ref{fig;combofits}(a)-(c). As before, the monoclinic model accurately captures the atomic PDF but not the unnormalized mPDF \dr, resulting in large difference curves.

We found that the atomic displacements allowed by the $C2$ symmetry of the nuclear structure refined to much smaller values than previously reported~\cite{goodw;prl06} and did not significantly improve the fits,
nor were they consistently reproducible from different starting parameter values. As such, we fixed these
displacements to zero. This raises the symmetry to $C2/m$, but for the sake of consistency,
we continue to use the $C2$ unit cell previously proposed~\cite{goodw;prl06}, but with all symmetry-lowering atomic displacements set to zero.

To check that the magnetic contributions to the PDF do not bias the structural
refinement if they are not removed before the refinement is carried out (the current {\it de-facto} situation in PDF refinements), we performed a second analysis in which we manually removed the visible magnetic Bragg peaks in reciprocal space
before computing the Fourier transform. The resulting parameter values are presented in  Table~\ref{table:mag-nomag}. The parameters refined to nearly identical
values as before, indicating that for small-box modeling the atomic PDF refinements are robust even with the
magnetic component present in the data but not included in the model.
\begin{table}
\centering 
\ra{1.3}
\caption{Refined parameters of the $C2$ model for nuclear PDF fits performed both with and without the magnetic Bragg peaks included in the Fourier transform. Values in parentheses represent the estimated standard deviations (ESDs) of the refined parameters. The ESDs of the parameters refined against the data without the magnetic Bragg peaks are less accurate due to errors introduced by manually removing the magnetic signal, but are expected to be on the same order of magnitude as indicated.} 

\begin{tabular}{c c c} 
\cline{1-3} 
 & With $I_m$ & Without $I_m$ \\  

\cline{1-3}

$a$ (\AA) & 5.4707(2) & 5.4681(8) \\
$b$ (\AA) & 3.1432(1) & 3.1443(7) \\
$c$ (\AA) & 15.1843(2) & 15.1819(8) \\
$\beta$ ($^{\circ}$) & 89.902(4) & 89.913(3) \\
$U_{Mn11}$ (\AA$^{-2}$) & 0.00302(5) & 0.00270(3) \\
$U_{Mn22}$ (\AA$^{-2}$) & 0.00124(5) & 0.00131(3) \\
$U_{Mn33}$ (\AA$^{-2}$) & 0.00193(5) & 0.00201(3) \\
$U_{O11}$ (\AA$^{-2}$) & 0.00282(4) & 0.00292(3) \\
$U_{O22}$ (\AA$^{-2}$) & 0.00284(6) & 0.00278(4) \\
$U_{O33}$ (\AA$^{-2}$) & 0.00316(4) & 0.00309(2) \\
$Q_{\textrm{damp}}$ (\AA$^{-1}$) & 0.0198(2) & 0.0138(1) \\
$Q_{\textrm{broad}}$ (\AA$^{-1}$) & 0.0195(5) & 0.0275(2) \\
$R_w$ (\%) & 24.8 & 6.5 \\

\cline{1-3}

\end{tabular}
\label{table:mag-nomag} 
\end{table}

Having completed a reliable atomic PDF fit, we then performed an mPDF fit to the experimental \dr\ contained in the atomic PDF difference curves. As before, we initially refined the model in which the overall
antiferromagnetic spin configuration was enforced as a constraint, but with the absolute direction of the spin axis allowed
to vary. Once again, the refined \dr\ very closely matched the experimental signal,
and the spin axis converged to lie in the (111) plane.

We next implemented the iterative fitting procedure described previously to further improve the quality of the fits. The initial fits were nearly as good as the final iterated fits, with the final $R_w$ typically representing an improvement of about 10\% of the original value. $R_w$ remained essentially unchanged after three iterations. The results of these atomic/magnetic PDF co-refinements are displayed in Fig.~\ref{fig;combofits}(d)-(f) for fitting ranges from 0 to 20~\AA, 60~\AA, and 100~\AA. The difference curves are much smaller than those for the atomic-only PDF fits in panels (a)-(c), and the final values of $R_w$ are between approximately 7\% and 10\%, indicating that these fits are of high quality. The results of this analysis therefore demonstrate that combined nuclear and magnetic neutron total scattering data can be used for quantitatively accurate co-refinements of atomic and magnetic structures.

\begin{figure}
\includegraphics[width=8.85cm]{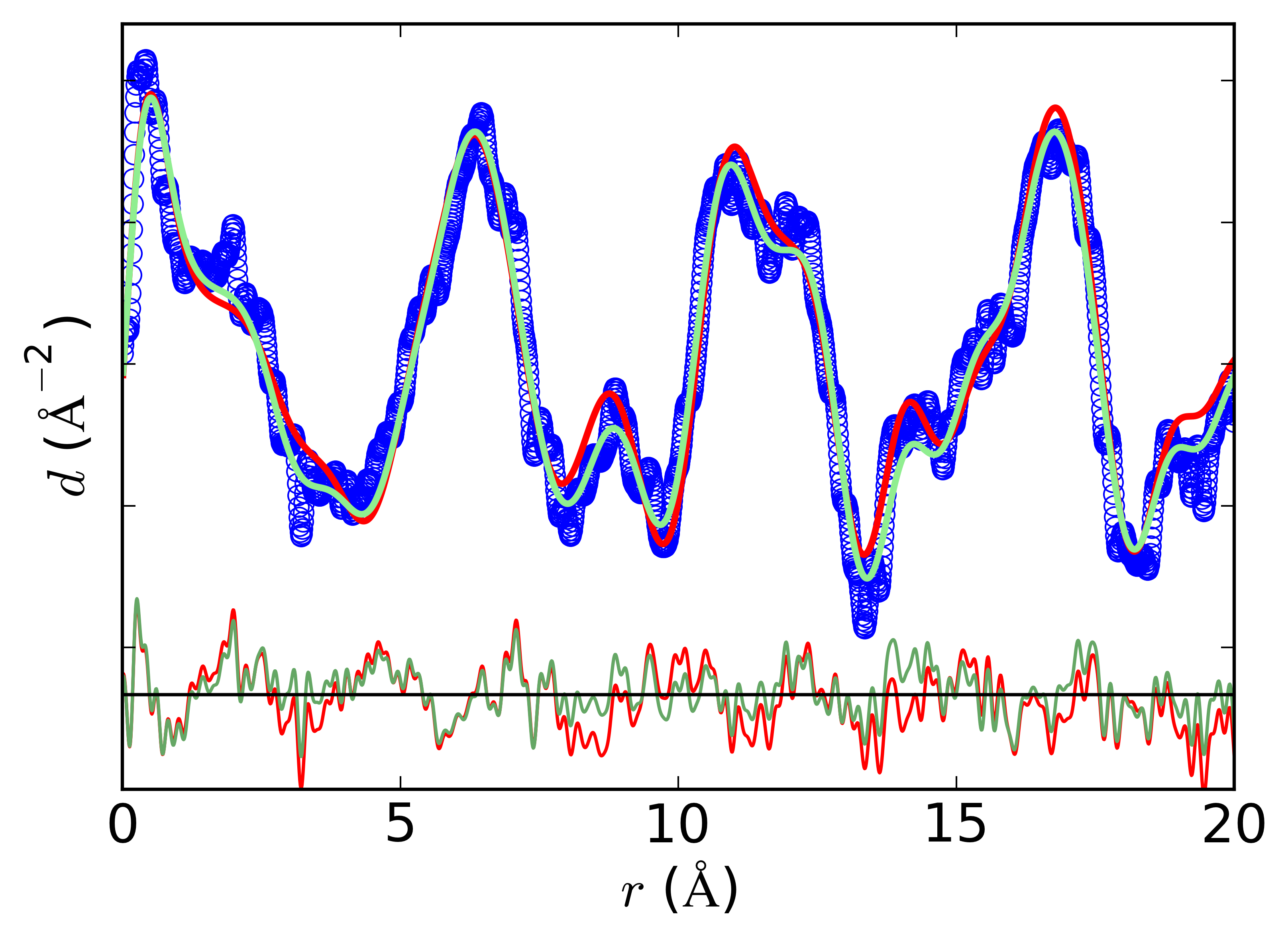}
\caption{Comparison of mPDF fits with the overall antiferromagnetic arrangement enforced as a constraint (red curve) and with all spin orientations within the monoclinic unit cell allowed to refine freely (green curve).}
\label{fig;free-v-constrained}
\end{figure}

Having verified the quantitative agreement between the constrained antiferromagnetic model and the experimental \dr, we next refined the unconstrained model described previously as an attempt at {\it ab initio} magnetic structure solution from mPDF analysis. The result of a 20~\AA\ fit is shown as the green curve in Fig.~\ref{fig;free-v-constrained}, along with the best fit of the highly constrained model described earlier shown in red for comparison.
The unconstrained fit is significantly better than the constrained fit when compared by eye, and the combined atomic/magnetic PDF fit value $R_w$ decreases from 6.88\% to 5.71\%. The freely refined spin configuration reproduces the expected overall antiferromagnetic structure, with spins closely aligned in common (111) sheets and anti-aligned between adjacent planes. These results indicate that magnetic structure solution is a realistic possibility for mPDF analysis.


As mentioned earlier, the monoclinic symmetry in principle allows for sensitivity to the direction of a preferred spin axis. The previous RMC study of MnO found a very slight preference for the pseudocubic $\langle 11\bar{2} \rangle$ directions~\cite{goodw;prl06}. To determine if the present data reveal a preferred spin axis, we performed approximately 50 unconstrained fits to the first 20~\AA\ of the experimental \dr. For a given refined spin configuration, we then summed the spin axes of all 12 spins in the unit cell to find the average axis for that configuration. These refinements converged robustly to a spin axis parallel to [10$\bar{1}$] to a high degree of precision, with $\hat{\boldsymbol{S}}\cdot(10\bar{1})/\sqrt{2}>0.99$ for all refinements, in contrast to the previous RMC findings~\cite{goodw;prl06}.

To investigate the reliability of these results, we performed an additional 50 unconstrained fits to data obtained by randomly subsampling the scattering data in reciprocal space at a sampling rate of 75\% before Fourier transforming into real space. For each data set, we refined the nuclear structure and then performed an unconstrained mPDF fit to the difference curve in the usual manner. The sub-sampling generates fifty distinct real-space data sets against which the model can be tested, allowing us to estimate not only a mean value for the preferred spin axis, but also a measure of the distribution of spin-axis directions.

\begin{figure}
\includegraphics[width=8.85cm]{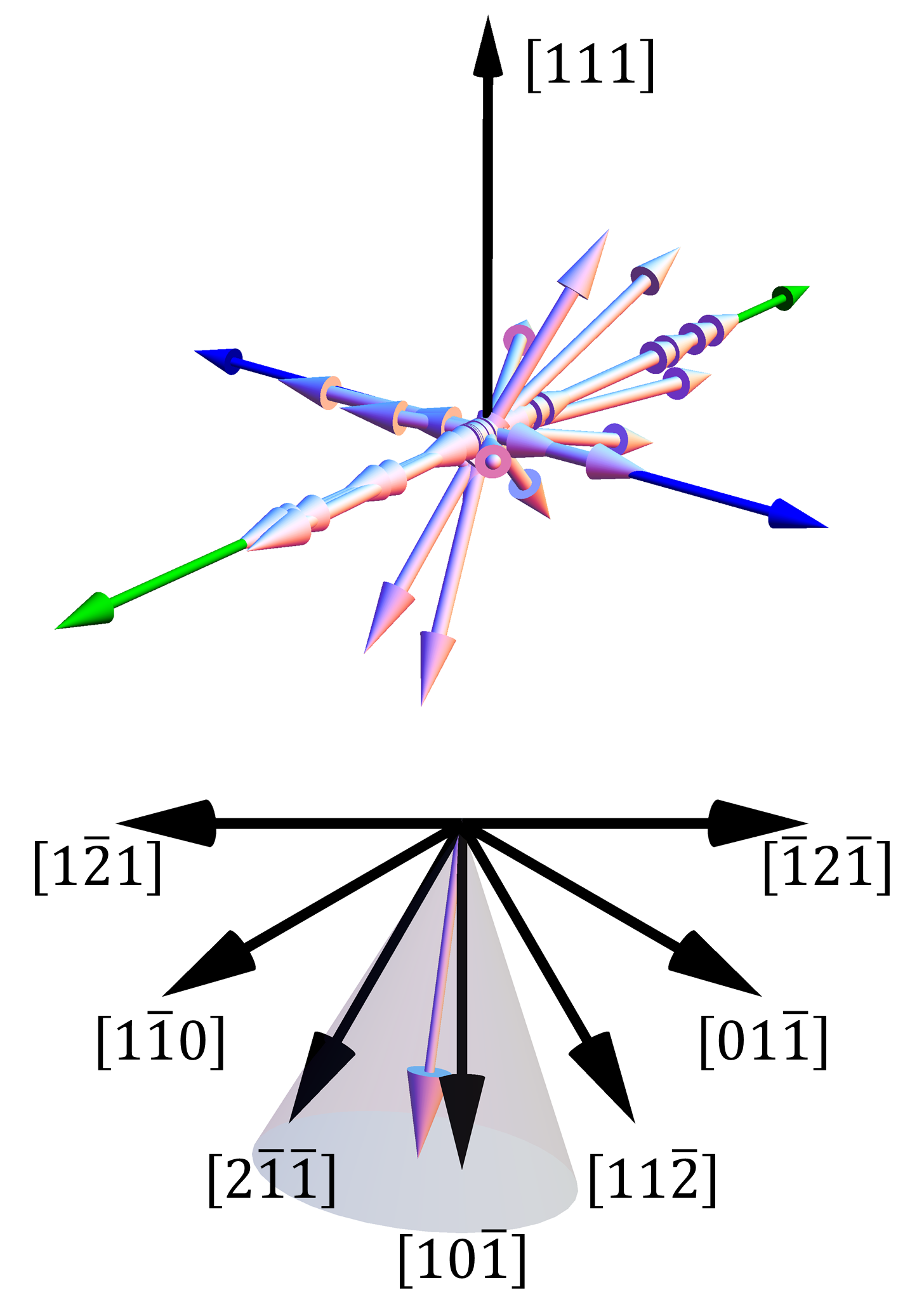}
\caption{Preferred spin axis in MnO. Top: Refined spin directions for 50 unconstrained fits performed using subsampled data, with the length of each arrow proportional to $\chi^{-2}$ for the corresponding refinement. The black arrow shows the [111] direction, and the green arrows show the [10$\bar{1}$] and symmetry-equivalent [$\bar{1}$01] directions, all in the pseudocubic setting. Blue arrows show the [$\bar{1}$2$\bar{1}$] and [1$\bar{2}$1] directions. Bottom: Projection of the weighted average of the symmetrized refined spin directions onto the (111) plane, with several high-symmetry directions shown as black arrows. The shaded region represents the width of the refined spin distribution ($\pm$ one standard deviation). }
\label{fig;subsampled-panel}
\end{figure}

The results of these 50 subsampled refinements are displayed in Fig.~\ref{fig;subsampled-panel}. The light-colored arrows in the top portion of the figure represent the configuration-averaged preferred spin directions for each individual refinement, weighted by $\chi^{-2}$ for the corresponding fit. It is clear that the refined spin directions are largely confined to the (111) plane and are clustered around the [10$\bar{1}$] and symmetry-equivalent [$\bar{1}$01] directions, indicated by long green arrows, albeit with some distribution. A smaller cluster of refined spin directions lies along the [1$\bar{2}$1] and [$\bar{1}$2$\bar{1}$]  directions. We then symmetrized all of the spin directions and computed the overall weighted average of the preferred spin direction. Transforming to a more convenient coordinate system in which the pseudocubic crystallographic [111] direction is the $z$-axis and [10$\bar{1}$] is the $x$-axis, we found an average polar angle of $\theta=94.4 \pm 18.1^{\circ}$ and an average azimuthal angle of $\phi=$-7.6 $\pm$ 28.7$^{\circ}$, where the magnitude of one standard deviation follows the $\pm$ sign. For reference, the high-symmetry direction [10$\bar{1}$] has polar and azimuthal angles of 90$^{\circ}$ and 0$^{\circ}$ in this coordinate system, while the neighboring [2$\bar{1}$$\bar{1}$] direction has $\theta=90^{\circ}$ and $\phi=30^{\circ}$. The bottom portion of Fig.~\ref{fig;subsampled-panel} displays the (111) plane with the overall weighted average of the refined spin directions shown as the light-colored arrow. The gray shaded region extends to $\pm$ one standard deviation of the refined azimuthal angle distribution. As can be seen from the figure, the preferred spin axis determined by these fits lies along [10$\bar{1}$] to within one standard deviation.


The unconstrained fits also allowed us to look for systematic out-of-plane canting of the spins. For nearly every refinement, most or all of the 12 spins in the unit cell possessed some small but nonzero out-of-plane component. The weighted average of the magnitude of the canting angle for all 50 subsampled refinements is $19.6 \pm 11.4^{\circ}$. This is consistent with the level of out-of-plane canting presented in the previous RMC study, although our results do not provide convincing evidence of the reported sinusoidal variation of the out-of-plane component~\cite{goodw;prl06}. Nevertheless, these results suggest that some form of non-collinearity of the magnetic moments may be present within the unit cell.

\subsection{Local versus average structure of MnO}

Finally, we utilized $r$-dependent atomic and magnetic PDF analysis to further investigate the relationship between the local monoclinic structure suggested by total scattering measurements and the average rhombohedral structure determined from high-resolution neutron diffraction. We performed atomic and magnetic PDF co-refinements over fitting ranges from 0 to 20~\AA, 60~\AA, and 100~\AA\ for both the
\R3m and $C2$ models. The results of the $r$-dependent co-refinements are displayed in Table~\ref{table:fits}.
\begin{table}
\centering 
\ra{1.3}
\caption{Refined parameters of combined atomic and magnetic PDF fits.} 

\begin{tabular}{c c c c c} 
\cline{1-5} 
 \multicolumn{2}{r}{Fitting Range:} & 0-20 \AA & 0-60 \AA & 0-100 \AA \\  

\cline{1-5}

\multirow{3}{*}{\R3m} & $R_w$ (\%) & 7.20 & 8.30 & 10.55 \\
& $a$ (\AA) & 3.15025(5) & 3.14948(2) & 3.14922(2) \\
& $c$ (\AA) & 7.5921(3) & 7.5915(1) & 7.5914(1) \\

\cline{1-5}

\multirow{5}{*}{$C2/m$}  & $R_w$ (\%) & 6.88 & 8.21 & 10.51 \\
& $a$ (\AA) & 5.4708(9) & 5.4609(4) & 5.4588(6) \\
& $b$ (\AA) & 3.1429(5) & 3.1460(4) & 3.1466(4) \\
& $c$ (\AA) & 15.1808(1) & 15.1825(2) & 15.1827(1) \\
& $\beta$ ($^{\circ}$) & 89.895(2) & 89.989(3) & 90.004(5) \\

\cline{1-5}

\end{tabular}
\label{table:fits} 
\end{table}
The overall $R_w$ values are slightly lower for the monoclinic model over all fitting ranges, supporting the scenario that the structure is indeed monoclinic on length scales up to 100~\AA. However, two observations suggest that the long-range average structure is actually approaching \R3m\ symmetry. First, the advantage in $R_w$ of the $C2$ model over \R3m decreases for the larger fitting ranges. Second, we note that
the $C2$ structure is equivalent to the \R3m structure when the monoclinic angle $\beta$ is 90$^{\circ}$ and the lattice parameter ratio $a/b$ is $\sqrt{3}\approx 1.732$. Inspection of Table~\ref{table:fits} shows that the ratio $a/b$ is 1.7407(4), 1.7358(3), and 1.7348(3) for the 20~\AA, 60~\AA, and 100~\AA\ fits, respectively,
with $\beta$ also approaching 90$^{\circ}$ for the higher fitting ranges. The PDF fits therefore suggest that the structure has monoclinic symmetry locally but tends toward rhombohedral symmetry on average. This may explain why only the rhombohedral structure has been reported in previous diffraction measurements, since they are sensitive only to the long-range, average structure and not the local structure. We suggest that misalignment of monoclinic domains on a length scale of approximately 100~\AA\ results in this average rhombohedral structure. The fact that these relatively large monoclinic domains do not result in clear peak splittings in the diffraction pattern points to the small magnitude of this monoclinic distortion. We suspect that a crystallographic refinement using the monoclinic structure may yield a better fit than the rhombohedral model by more accurately capturing the broadening of the peaks due to the small monoclinic distortion.

\section{Summary}

Neutron scattering data have been collected and analyzed from MnO as the first experimental application of the recently derived mPDF equations. A practical method is presented for performing simultaneous structural and magnetic PDF co-refinements using the unnormalized mPDF function \dr. The results suggest the presence of local monoclinic symmetry in MnO with a preferred spin alignment axis of [10$\bar{1}$]. The method of co-refining structural and magnetic PDFs obtained from unpolarized neutrons appears to be a promising method for studies of magnetic materials.


\section{Appendix: Derivation of the unnormalized mPDF function \dr}
As shown in Eq.~\ref{totalPDF}, the unnormalized mPDF is given by $d(r)=\frac{2}{\pi}\int_{0}^{\infty} QI_m\sin{Q r}\text{d}Q$ and is related to the normalized mPDF $\mathpzc{f}(r)$ as in Eq.~\ref{eq;dr}. Here we provide the derivation of Eq.~\ref{eq;dr} and explicitly work out the constants $C_1$ and $C_2$. We first define the constant $K=\frac{2}{3}S(S+1)(\gamma r_0)^2$ and rewrite Eq.~\ref{FT} as
\begin{align}
\mathpzc{f}(r)=\frac{2}{\pi}\int_{0}^{\infty} Q\left(\frac{I_m}{N_s K f_{m}^{2}}-1\right)\sin{Q r}\text{d}Q \label{compactFT}.
\end{align}
We have suppressed the explicit $Q$-dependence of the magnetic form factor $f_{m}$ for notational compactness. We now write \dr in the form
\begin{align}
d(r)&=\frac{2}{\pi}\int_{0}^{\infty} Q\left[\left(\frac{I_m}{N_s K f_{m}^2}-1\right)N_s K f_{m}^2 + N_s K f_{m}^2\right]\sin{Q r}\text{d}Q \label{dr2}
\\&=N_s K \frac{2}{\pi}\int_{0}^{\infty} Q\left(\frac{I_m}{N_s K f_{m}^2}-1\right)f_{m}^2\sin{Q r}\text{d}Q + N_s K \frac{2}{\pi}\int_{0}^{\infty} Qf_{m}^2\sin{Q r}\text{d}Q\label{drexpanded}.
\end{align}
We will employ the convolution theorem to relate Eq.~\ref{drexpanded} to $\mathpzc{f}(r)$. Defining the Fourier transform of a function $H(Q)$ as $\mathcal{F}\left\{  H(Q)\right\} =\frac{1}{\sqrt{2\pi}}\int_{-\infty}^{\infty}H(Q)\exp i Qr \text{d}Q$ and the convolution of two functions $h(r)$ and $j(r)$ as $h\ast r = \int_{-\infty}^{\infty}h(r')j(r-r')\text{d}r'$, the convolution theorem states that $\mathcal{F}\left\{  H(Q)J(Q)\right\} =\frac{1}{\sqrt{2\pi}}\mathcal{F}\left\{  H(Q)\right\} \ast \mathcal{F}\left\{  J(Q)\right\} .$
To apply this to the sine transforms used in the definitions of \dr and $\mathpzc{f}(r)$, we note that for an odd function $g(Q)$,
\begin{align}
\frac{2}{\pi}\int_{0}^{\infty}g(Q)\sin{Qr}\text{d}Q = -i \sqrt{\frac{2}{\pi}}\mathcal{F}\left\{ g(Q)\right\} .
\end{align}
Therefore, \dr can be written as
\begin{align}
d(r)=-i N_s K \sqrt{\frac{2}{\pi}}\left( \mathcal{F}\left\{ Q\left( \frac{I_m}{N_s K f_{m}^2}-1 \right)f_{m}^2\right\}  +\mathcal{F}\left\{  Qf_{m}^2 \right\} \right)\label{drFT}.
\end{align}
Applying the convolution theorem to each term in Eq.~\ref{drFT}, we have
\begin{align}
\mathcal{F}\left\{ Q\left( \frac{I_m}{N_s K f_{m}^2}-1 \right)f_{m}^2\right\} &=\frac{1}{\sqrt{2\pi}}\mathcal{F}\left\{  Q\left( \frac{I_m}{N_s K f_{m}^2}-1 \right)\right\}  \ast  \left( \frac{1}{\sqrt{2\pi}} \mathcal{F}\left\{ f\right\}  \ast \mathcal{F}\left\{ f\right\}   \right)
\\&=\frac{1}{2\pi}\mathcal{F}\left\{  Q\left( \frac{I_m}{N_s K f_{m}^2}-1 \right)\right\}  \ast S(r) \label{drFTterm1}
\end{align}
and
\begin{align}
\mathcal{F}\left\{ Qf_{m}^2\right\} &=\frac{1}{2\pi}\mathcal{F}\left\{  Q \right\}  \ast \mathcal{F}\left\{  f \right\}  \ast \mathcal{F}\left\{  f \right\}
\\&=\frac{1}{2\pi}\mathcal{F}\left\{  Q \right\}  \ast S(r),
\end{align}
where $S(r)=\mathcal{F}\left\{ f(q)\right\} \ast\mathcal{F}\left\{ f(q)\right\} $ as described in the main text. Noting that $\mathcal{F}\left\{ Q\right\} =\frac{1}{\sqrt{2\pi}}\int_{-\infty}^{\infty}Q\exp{i Qr}\text{d}Q=-i \sqrt{2\pi}\delta '(r)$, and using the properties of the delta function, we have
\begin{align}
\mathcal{F}\left\{ Qf_{m}^2\right\} &=\frac{1}{2\pi}\left( -i \sqrt{2\pi}\delta ' (r) \ast S(r) \right)
\\&=\frac{-i}{\sqrt{2\pi}}S'(r).
\end{align}
Substituting these results back into Eq.~\ref{drFT} yields
\begin{align}
d(r)&=\frac{N_s K}{2\pi}\left( -i \sqrt{\frac{2}{\pi}} \mathcal{F}\left\{  Q\left( \frac{I_m}{N_s K f_{m}^2}-1 \right) \right\}  \ast S(r) \right) - \frac{N_s K}{\sqrt{2\pi}}S'(r)
\\&=\frac{N_s K}{2\pi} \frac{2}{\pi}\int\limits_{0}^{\infty}Q\left( \frac{I_m}{N_s K f_{m}^2}-1\right)\sin{Qr}\text{d}Q \ast S(r) - \frac{N_s K}{\sqrt{2\pi}}S'(r)
\\&=\frac{N_s K}{2\pi}\mathpzc{f}(r)\ast S(r)-\frac{N_s K}{\sqrt{2\pi}}S'(r)\label{finaldr}.
\end{align}
This result defines the constants $C_1$ and $C_2$ in Eq.~\ref{eq;dr}. We note that the quantity $S'(r)$ is generally negative for $r>0$, so the second term in Eq.~\ref{finaldr} gives rise to the positive peak at low $r$ evident in the experimental unnormalized mPDFs. Finally, we point out that the ratio of the two scaling constants $C_1$ and $C_2$ in Eq.~\ref{eq;dr} is $\lvert C_1 / C_2 \rvert = 1 / \sqrt{2\pi}\approx 0.399$. In the refinements described in the text, both $C_1$ and $C_2$ were allowed to vary independently. However, the ratio of the refined values was always within 3\% of the predicted ratio of $1 / \sqrt{2\pi}$, thereby verifying this relationship.


\textbf{Acknowledgements}
We gratefully acknowledge Michela Brunelli at the ILL and Joan Siewenie at the Lujan Center for valuable assistance during the data collection and processing stages, as well as Kate Page at the Lujan Center and Xiaohao Yang at Columbia University for useful discussions. BF was supported by the U.S. National Science Foundation (NSF) Partnership in International Research and Education initiative (PIRE) via Grant No. PIRE: OISE-0968226 and by the NSF Graduate Research Fellowship via Grant No. DGE --- 11-44155, and SJLB was supported by the U.S. Department of Energy, Office of Basic Energy Sciences, under Contract No. DE-AC02-98CH10886. Neutron scattering experiments were carried out on NPDF at LANSCE, funded by DOE Office of Basic Energy Sciences. LANL is operated by Los Alamos National Security LLC under DOE Contract No. DE-AC52-06NA25396.



\begin{thebibliography}{19}%
\makeatletter
\providecommand \@ifxundefined [1]{%
 \@ifx{#1\undefined}
}%
\providecommand \@ifnum [1]{%
 \ifnum #1\expandafter \@firstoftwo
 \else \expandafter \@secondoftwo
 \fi
}%
\providecommand \@ifx [1]{%
 \ifx #1\expandafter \@firstoftwo
 \else \expandafter \@secondoftwo
 \fi
}%
\providecommand \natexlab [1]{#1}%
\providecommand \enquote  [1]{``#1''}%
\providecommand \bibnamefont  [1]{#1}%
\providecommand \bibfnamefont [1]{#1}%
\providecommand \citenamefont [1]{#1}%
\providecommand \href@noop [0]{\@secondoftwo}%
\providecommand \href [0]{\begingroup \@sanitize@url \@href}%
\providecommand \@href[1]{\@@startlink{#1}\@@href}%
\providecommand \@@href[1]{\endgroup#1\@@endlink}%
\providecommand \@sanitize@url [0]{\catcode `\\12\catcode `\$12\catcode
  `\&12\catcode `\#12\catcode `\^12\catcode `\_12\catcode `\%12\relax}%
\providecommand \@@startlink[1]{}%
\providecommand \@@endlink[0]{}%
\providecommand \url  [0]{\begingroup\@sanitize@url \@url }%
\providecommand \@url [1]{\endgroup\@href {#1}{\urlprefix }}%
\providecommand \urlprefix  [0]{URL }%
\providecommand \Eprint [0]{\href }%
\providecommand \doibase [0]{http://dx.doi.org/}%
\providecommand \selectlanguage [0]{\@gobble}%
\providecommand \bibinfo  [0]{\@secondoftwo}%
\providecommand \bibfield  [0]{\@secondoftwo}%
\providecommand \translation [1]{[#1]}%
\providecommand \BibitemOpen [0]{}%
\providecommand \bibitemStop [0]{}%
\providecommand \bibitemNoStop [0]{.\EOS\space}%
\providecommand \EOS [0]{\spacefactor3000\relax}%
\providecommand \BibitemShut  [1]{\csname bibitem#1\endcsname}%
\let\auto@bib@innerbib\@empty
\bibitem [{\citenamefont {Frandsen}\ \emph {et~al.}(2014)\citenamefont
  {Frandsen}, \citenamefont {Yang},\ and\ \citenamefont
  {Billinge}}]{frand;aca14}%
  \BibitemOpen
  \bibfield  {author} {\bibinfo {author} {\bibfnamefont {B.~A.}\ \bibnamefont
  {Frandsen}}, \bibinfo {author} {\bibfnamefont {X.}~\bibnamefont {Yang}}, \
  and\ \bibinfo {author} {\bibfnamefont {S.~J.}\ \bibnamefont {Billinge}},\
  }\href {\doibase 10.1107/S2053273313033081} {\bibfield  {journal} {\bibinfo
  {journal} {Acta Crystallogr. A}\ }\textbf {\bibinfo {volume} {70}},\ \bibinfo
  {pages} {3} (\bibinfo {year} {2014})}\BibitemShut {NoStop}%
\bibitem [{\citenamefont {Egami}\ and\ \citenamefont
  {Billinge}(2012)}]{egami;b;utbp12}%
  \BibitemOpen
  \bibfield  {author} {\bibinfo {author} {\bibfnamefont {T.}~\bibnamefont
  {Egami}}\ and\ \bibinfo {author} {\bibfnamefont {S.~J.~L.}\ \bibnamefont
  {Billinge}},\ }\href@noop {} {\emph {\bibinfo {title} {Underneath the Bragg
  peaks: structural analysis of complex materials}}},\ \bibinfo {edition}
  {2nd}\ ed.\ (\bibinfo  {publisher} {Elsevier},\ \bibinfo {address}
  {Amsterdam},\ \bibinfo {year} {2012})\BibitemShut {NoStop}%
\bibitem [{\citenamefont {Shull}\ \emph {et~al.}(1951)\citenamefont {Shull},
  \citenamefont {Strauser},\ and\ \citenamefont {Wollan}}]{shull;pr51}%
  \BibitemOpen
  \bibfield  {author} {\bibinfo {author} {\bibfnamefont {C.~G.}\ \bibnamefont
  {Shull}}, \bibinfo {author} {\bibfnamefont {W.~A.}\ \bibnamefont {Strauser}},
  \ and\ \bibinfo {author} {\bibfnamefont {E.~O.}\ \bibnamefont {Wollan}},\
  }\href@noop {} {\bibfield  {journal} {\bibinfo  {journal} {Phys. Rev.}\
  }\textbf {\bibinfo {volume} {83}},\ \bibinfo {pages} {333} (\bibinfo {year}
  {1951})}\BibitemShut {NoStop}%
\bibitem [{\citenamefont {Roth}(1958)}]{roth;pr58}%
  \BibitemOpen
  \bibfield  {author} {\bibinfo {author} {\bibfnamefont {W.~L.}\ \bibnamefont
  {Roth}},\ }\href@noop {} {\bibfield  {journal} {\bibinfo  {journal} {Phys.
  Rev.}\ }\textbf {\bibinfo {volume} {110}},\ \bibinfo {pages} {1333} (\bibinfo
  {year} {1958})}\BibitemShut {NoStop}%
\bibitem [{\citenamefont {Goodwin}\ \emph {et~al.}(2006)\citenamefont
  {Goodwin}, \citenamefont {Tucker}, \citenamefont {Dove},\ and\ \citenamefont
  {Keen}}]{goodw;prl06}%
  \BibitemOpen
  \bibfield  {author} {\bibinfo {author} {\bibfnamefont {A.~L.}\ \bibnamefont
  {Goodwin}}, \bibinfo {author} {\bibfnamefont {M.~G.}\ \bibnamefont {Tucker}},
  \bibinfo {author} {\bibfnamefont {M.~T.}\ \bibnamefont {Dove}}, \ and\
  \bibinfo {author} {\bibfnamefont {D.~A.}\ \bibnamefont {Keen}},\ }\href
  {\doibase 10.1103/PhysRevLett.96.047209} {\bibfield  {journal} {\bibinfo
  {journal} {Phys. Rev. Lett.}\ }\textbf {\bibinfo {volume} {96}},\ \bibinfo
  {pages} {047209} (\bibinfo {year} {2006})}\BibitemShut {NoStop}%
\bibitem [{\citenamefont {Shaked}\ \emph {et~al.}(1988)\citenamefont {Shaked},
  \citenamefont {Jaber},\ and\ \citenamefont {Hitterman}}]{shake;prb88}%
  \BibitemOpen
  \bibfield  {author} {\bibinfo {author} {\bibfnamefont {H.}~\bibnamefont
  {Shaked}}, \bibinfo {author} {\bibfnamefont {J.~J.}\ \bibnamefont {Jaber}}, \
  and\ \bibinfo {author} {\bibfnamefont {R.}~\bibnamefont {Hitterman}},\
  }\href@noop {} {\bibfield  {journal} {\bibinfo  {journal} {Phys. Rev. B}\
  }\textbf {\bibinfo {volume} {38}},\ \bibinfo {pages} {11901} (\bibinfo {year}
  {1988})}\BibitemShut {NoStop}%
\bibitem [{\citenamefont {Mellerg{\aa}rd}\ \emph {et~al.}(1998)\citenamefont
  {Mellerg{\aa}rd}, \citenamefont {McGreevy}, \citenamefont {Wannberg},\ and\
  \citenamefont {Trostell}}]{melle;jpcm98}%
  \BibitemOpen
  \bibfield  {author} {\bibinfo {author} {\bibfnamefont {A.}~\bibnamefont
  {Mellerg{\aa}rd}}, \bibinfo {author} {\bibfnamefont {R.}~\bibnamefont
  {McGreevy}}, \bibinfo {author} {\bibfnamefont {A.}~\bibnamefont {Wannberg}},
  \ and\ \bibinfo {author} {\bibfnamefont {B.}~\bibnamefont {Trostell}},\
  }\href@noop {} {\bibfield  {journal} {\bibinfo  {journal} {J. Phys.: Condens.
  Mat.}\ }\textbf {\bibinfo {volume} {10}},\ \bibinfo {pages} {9401} (\bibinfo
  {year} {1998})}\BibitemShut {NoStop}%
\bibitem [{\citenamefont {McGreevy}(2001)}]{mcgre;jpcm01}%
  \BibitemOpen
  \bibfield  {author} {\bibinfo {author} {\bibfnamefont {R.~L.}\ \bibnamefont
  {McGreevy}},\ }\href {\doibase 10.1088/0953-8984/13/46/201} {\bibfield
  {journal} {\bibinfo  {journal} {J. Phys.: Condens. Mat.}\ }\textbf {\bibinfo
  {volume} {13}},\ \bibinfo {pages} {R877} (\bibinfo {year}
  {2001})}\BibitemShut {NoStop}%
\bibitem [{\citenamefont {Paddison}\ and\ \citenamefont
  {Goodwin}(2012)}]{goodw;prl12}%
  \BibitemOpen
  \bibfield  {author} {\bibinfo {author} {\bibfnamefont {J.~A.~M.}\
  \bibnamefont {Paddison}}\ and\ \bibinfo {author} {\bibfnamefont {A.~L.}\
  \bibnamefont {Goodwin}},\ }\href {\doibase 10.1103/PhysRevLett.108.017204}
  {\bibfield  {journal} {\bibinfo  {journal} {Phys. Rev. Lett.}\ }\textbf
  {\bibinfo {volume} {108}},\ \bibinfo {pages} {017204} (\bibinfo {year}
  {2012})}\BibitemShut {NoStop}%
\bibitem [{\citenamefont {Blech}\ and\ \citenamefont
  {Averbach}(1964)}]{blech;p64}%
  \BibitemOpen
  \bibfield  {author} {\bibinfo {author} {\bibfnamefont {I.~A.}\ \bibnamefont
  {Blech}}\ and\ \bibinfo {author} {\bibfnamefont {B.~L.}\ \bibnamefont
  {Averbach}},\ }\href@noop {} {\bibfield  {journal} {\bibinfo  {journal}
  {Phys.}\ }\textbf {\bibinfo {volume} {1}},\ \bibinfo {pages} {31} (\bibinfo
  {year} {1964})}\BibitemShut {NoStop}%
\bibitem [{\citenamefont {Farrow}\ and\ \citenamefont
  {Billinge}(2009)}]{farro;aca09}%
  \BibitemOpen
  \bibfield  {author} {\bibinfo {author} {\bibfnamefont {C.~L.}\ \bibnamefont
  {Farrow}}\ and\ \bibinfo {author} {\bibfnamefont {S.~J.~L.}\ \bibnamefont
  {Billinge}},\ }\href@noop {} {\bibfield  {journal} {\bibinfo  {journal} {Acta
  Crystallogr. A}\ }\textbf {\bibinfo {volume} {65}},\ \bibinfo {pages} {232}
  (\bibinfo {year} {2009})}\BibitemShut {NoStop}%
\bibitem [{\citenamefont {Rodriguez-Carvajal}(1990)}]{rodri;unpub90}%
  \BibitemOpen
  \bibfield  {author} {\bibinfo {author} {\bibfnamefont {J.}~\bibnamefont
  {Rodriguez-Carvajal}},\ }in\ \href@noop {} {\emph {\bibinfo {booktitle}
  {Abstracts of the Satellite Meeting on Powder Diffraction of the XV Congress
  of the IUCr, Toulouse, France}}}\ (\bibinfo {year} {1990})\BibitemShut
  {NoStop}%
\bibitem [{\citenamefont {Wilson}(1995)}]{wilso;b;itc95}%
  \BibitemOpen
  \bibinfo {editor} {\bibfnamefont {A.}~\bibnamefont {Wilson}},\ ed.,\
  \href@noop {} {\emph {\bibinfo {title} {International Tables for
  Crystallography, Volume C: Mathematical, Physical, and Chemical Tables}}}\
  (\bibinfo  {publisher} {Kluwer},\ \bibinfo {address} {Dordrecht, The
  Netherlands},\ \bibinfo {year} {1995})\BibitemShut {NoStop}%
\bibitem [{\citenamefont {Stewart}\ \emph {et~al.}(2009)\citenamefont
  {Stewart}, \citenamefont {Deen}, \citenamefont {Andersen}, \citenamefont
  {Schober}, \citenamefont {Barthelemy}, \citenamefont {Hillier}, \citenamefont
  {Murani}, \citenamefont {Hayes},\ and\ \citenamefont
  {Lindenau}}]{stewa;jac09}%
  \BibitemOpen
  \bibfield  {author} {\bibinfo {author} {\bibfnamefont {J.~R.}\ \bibnamefont
  {Stewart}}, \bibinfo {author} {\bibfnamefont {P.~P.}\ \bibnamefont {Deen}},
  \bibinfo {author} {\bibfnamefont {K.~H.}\ \bibnamefont {Andersen}}, \bibinfo
  {author} {\bibfnamefont {H.}~\bibnamefont {Schober}}, \bibinfo {author}
  {\bibfnamefont {J.-F.}\ \bibnamefont {Barthelemy}}, \bibinfo {author}
  {\bibfnamefont {J.~M.}\ \bibnamefont {Hillier}}, \bibinfo {author}
  {\bibfnamefont {A.~P.}\ \bibnamefont {Murani}}, \bibinfo {author}
  {\bibfnamefont {T.}~\bibnamefont {Hayes}}, \ and\ \bibinfo {author}
  {\bibfnamefont {B.}~\bibnamefont {Lindenau}},\ }\href@noop {} {\bibfield
  {journal} {\bibinfo  {journal} {J. Appl. Crystallogr.}\ }\textbf {\bibinfo
  {volume} {42}},\ \bibinfo {pages} {69} (\bibinfo {year} {2009})}\BibitemShut
  {NoStop}%
\bibitem [{\citenamefont {Schweika}(2010)}]{schwe;jpconfs10}%
  \BibitemOpen
  \bibfield  {author} {\bibinfo {author} {\bibfnamefont {W.}~\bibnamefont
  {Schweika}},\ }\href@noop {} {\bibfield  {journal} {\bibinfo  {journal} {J.
  Phys.: Conf. Ser.}\ }\textbf {\bibinfo {volume} {211}},\ \bibinfo {pages}
  {012026} (\bibinfo {year} {2010})}\BibitemShut {NoStop}%
\bibitem [{\citenamefont {Ehlers}\ \emph {et~al.}(2013)\citenamefont {Ehlers},
  \citenamefont {Stewart}, \citenamefont {Wildes}, \citenamefont {Deen},\ and\
  \citenamefont {Andersen}}]{ehler;rsi13}%
  \BibitemOpen
  \bibfield  {author} {\bibinfo {author} {\bibfnamefont {G.}~\bibnamefont
  {Ehlers}}, \bibinfo {author} {\bibfnamefont {J.~R.}\ \bibnamefont {Stewart}},
  \bibinfo {author} {\bibfnamefont {A.~R.}\ \bibnamefont {Wildes}}, \bibinfo
  {author} {\bibfnamefont {P.~P.}\ \bibnamefont {Deen}}, \ and\ \bibinfo
  {author} {\bibfnamefont {K.~H.}\ \bibnamefont {Andersen}},\ }\href@noop {}
  {\bibfield  {journal} {\bibinfo  {journal} {Rev. Sci. Instrum.}\ }\textbf
  {\bibinfo {volume} {84}},\ \bibinfo {pages} {093901} (\bibinfo {year}
  {2013})}\BibitemShut {NoStop}%
\bibitem [{\citenamefont {Peterson}\ \emph {et~al.}(2000)\citenamefont
  {Peterson}, \citenamefont {Gutmann}, \citenamefont {Proffen},\ and\
  \citenamefont {Billinge}}]{peter;jac00}%
  \BibitemOpen
  \bibfield  {author} {\bibinfo {author} {\bibfnamefont {P.~F.}\ \bibnamefont
  {Peterson}}, \bibinfo {author} {\bibfnamefont {M.}~\bibnamefont {Gutmann}},
  \bibinfo {author} {\bibfnamefont {T.}~\bibnamefont {Proffen}}, \ and\
  \bibinfo {author} {\bibfnamefont {S.~J.~L.}\ \bibnamefont {Billinge}},\
  }\href@noop {} {\bibfield  {journal} {\bibinfo  {journal} {J. Appl.
  Crystallogr.}\ }\textbf {\bibinfo {volume} {33}},\ \bibinfo {pages} {1192}
  (\bibinfo {year} {2000})}\BibitemShut {NoStop}%
\bibitem [{\citenamefont {Farrow}\ \emph {et~al.}(2007)\citenamefont {Farrow},
  \citenamefont {Juh\'as}, \citenamefont {Liu}, \citenamefont {Bryndin},
  \citenamefont {{Bo\v zin}}, \citenamefont {Bloch}, \citenamefont {Proffen},\
  and\ \citenamefont {Billinge}}]{farro;jpcm07}%
  \BibitemOpen
  \bibfield  {author} {\bibinfo {author} {\bibfnamefont {C.~L.}\ \bibnamefont
  {Farrow}}, \bibinfo {author} {\bibfnamefont {P.}~\bibnamefont {Juh\'as}},
  \bibinfo {author} {\bibfnamefont {J.}~\bibnamefont {Liu}}, \bibinfo {author}
  {\bibfnamefont {D.}~\bibnamefont {Bryndin}}, \bibinfo {author} {\bibfnamefont
  {E.~S.}\ \bibnamefont {{Bo\v zin}}}, \bibinfo {author} {\bibfnamefont
  {J.}~\bibnamefont {Bloch}}, \bibinfo {author} {\bibfnamefont
  {T.}~\bibnamefont {Proffen}}, \ and\ \bibinfo {author} {\bibfnamefont
  {S.~J.~L.}\ \bibnamefont {Billinge}},\ }\href {\doibase
  10.1088/0953-8984/19/33/335219} {\bibfield  {journal} {\bibinfo  {journal}
  {J. Phys.: Condens. Mat.}\ }\textbf {\bibinfo {volume} {19}},\ \bibinfo
  {pages} {335219} (\bibinfo {year} {2007})}\BibitemShut {NoStop}%
\bibitem [{\citenamefont {Peterson}\ \emph {et~al.}(2003)\citenamefont
  {Peterson}, \citenamefont {{E. S. Bo\v zin}}, \citenamefont {Proffen},\ and\
  \citenamefont {Billinge}}]{peter;jac03}%
  \BibitemOpen
  \bibfield  {author} {\bibinfo {author} {\bibfnamefont {P.~F.}\ \bibnamefont
  {Peterson}}, \bibinfo {author} {\bibnamefont {{E. S. Bo\v zin}}}, \bibinfo
  {author} {\bibfnamefont {T.}~\bibnamefont {Proffen}}, \ and\ \bibinfo
  {author} {\bibfnamefont {S.~J.~L.}\ \bibnamefont {Billinge}},\ }\href@noop {}
  {\bibfield  {journal} {\bibinfo  {journal} {J. Appl. Crystallogr.}\ }\textbf
  {\bibinfo {volume} {36}},\ \bibinfo {pages} {53} (\bibinfo {year}
  {2003})}\BibitemShut {NoStop}%
\end{thebibliography}
%

\end{document}